\documentclass[twocolumn,prl,10pt,aps,floatfix]{revtex4-1}
\usepackage{graphicx}
\usepackage{amsmath}
\usepackage[colorlinks,linkcolor=blue,citecolor=blue,urlcolor=blue]{hyperref}
\usepackage{color}

\begin{document}
\title{A note on the deformation of 1D ferromagnetic 
domain walls\\
due to double exchange interaction with a free electron gas}
\author{K. Seremetas and X. Zotos}
\affiliation{Department of Physics,
University of Crete, 70013 Heraklion, Greece}
\date{\today}

\begin{abstract}
Using an S-matrix formulation we evaluate the thermodynamic potential 
and conductance
of a Bloch or N\'eel  magnetic wall interacting with an open one dimensional 
free electron gas via a double exchange interaction. 
The minimization of the elastic magnetic energy plus electronic 
thermodynamic potential 
indicates that for chemical potential larger than the magnetic 
interaction the domain wall is generally deformed to a thin wall,  
while for magnetic interaction larger than the chemical potential 
tends towards wide walls.
In contrast, for a double wall configuration the deformation 
is always towards wide walls.
For the single as well as the double magnetic wall configurations the  
conductance monotonically decreases with decreasing wall width.
The thermodynamic potential 
and conductance of Bloch and N\'eel magnetic domain walls are 
identical within this prototype model.
\end{abstract}
\maketitle

\section{Introduction}
Magnetic topological textures as for instance Bloch, N\'eel domain walls and 
skyrmions, play a fundamental role in the field of magnonics \cite{roadmap}, 
spin-wave computing and skyrmionics.
The study of magnetic domain walls is an  extremely 
extensive field 
both in fundamental science \cite{topology} and engineering \cite{apps}.

On the bulk level, metallic magnetic systems have been described 
by the so-called double exchange model which 
describes a lattice of classical spins interacting with the 
conduction electrons through the Hund's rule coupling which 
aligns the spins of the conduction band and localized electrons 
occupying the same lattice site. 
The double exchange interaction was shown to lead to complex magnetic phases 
\cite{brink,mostovoy}.
Furthermore, the experimental issue of conductance and shape of magnetic 
textures e.g. in manganites \cite{mathur} or
the deformation of a skyrmion by an electric current \cite{yasin}
has been previously studied. Theoretically, the conductance 
\cite{fukuyama,levy,nagaosa} 
and the ferromagnetic wall deformation in one and   
two-dimensional systems have been discussed in 
\cite{tatara,golosov1,golosov2,motome}.

In this work we study a generic model of a one dimensional classical magnetic system
with Bloch/N\'eel domain
walls interacting via a double exchange interaction with a free electron system.
Thus we neglect the quantum nature of the localized spins, an approximation
appropriate for large spins.
We should emphasize that we are dealing with an
open quantum mechanical coherent system.
We use the S-matrix formulation \cite{s_matrix,buttiker,jellium} 
of statistical mechanics for an open system
that provides a rigorous and unified framework
for the evaluation of the thermodynamic potential 
and at the same time the conductance of the electronic gas.
The Landauer formulation we are using implies ballistic character of 
electronic transport, thus relevant to mesoscopic systems.
Single chain molecular magnets \cite{single1,single2} or single chain magnets
on a metallic substrate could be candidates for the following analysis.

Besides this prototype one dimensional model, the method we are using 
can be extended to the study of the 
deformation of higher dimensional magnetic textures as skyrmions 
due to the double exchange interaction with an electronic system. 
It can also be incorporated in numerical methods \cite{kwant} 
for the study of quantum transport in magnetic structures. 

\section{Model and method}
We consider a one dimensional classical continuous magnetic system 
of length $L$ described by the energy,
\begin{equation}
E_m=J\int_{-L/2}^{L/2} dx {\Big(}\frac{\partial \theta}{\partial x}{\Big )}^2
+D\int_{-L/2}^{L/2} dx \sin^2 \theta(x),
\label{ew}
\end{equation}
\noindent
where $\theta(x)$ is the angle of the magnetic moment from the z-axis, 
$J$ the exchange and $D$ the anisotropy interaction.
For a single magnetic wall, 
minimizing the energy with boundary conditions 
$\theta \rightarrow \pi$ for $x\rightarrow -L/2$ 
and $\theta \rightarrow 0$ for $x\rightarrow +L/2$, we obtain the
Bloch domain wall  profile,
\begin{equation}
\theta(x)=2\tan^{-1} e^{-x/\xi}
\label{profile}
\end{equation}

\noindent
of energy $E_m=2J/\xi+2D\xi$ and  width at minimum energy, 
$\xi_m=\sqrt{\frac{J}{D}}$, for $L\rightarrow \infty$.

The magnetic system interacts with an open one dimensional 
bath of free electrons described by the equation,

\begin{equation}
\Big[-\frac{\hbar^2}{2m}\frac{\partial^2}{\partial x^2} +V(x)\Big]\Psi
=\epsilon\Psi
\end{equation}

\noindent
where $\Psi$ 
is a two component plane wave wavefunction of wavevector $q$ 
for the z-projection of the electron spin. 
$V(x)$ is the double exchange interaction,

\begin{eqnarray}
V(x)&=&-\vec {h}(x) \cdot \vec \sigma=-h_x\sigma^x-h_z\sigma^z
\nonumber\\
&=&-h
\begin{pmatrix}
+\cos\theta(x) & +\sin\theta(x)\\
+\sin\theta(x) & -\cos \theta(x)
\end{pmatrix}.
\end{eqnarray}

\noindent
The fictitious magnetic field $h$ is a product of the magnetic 
system spin and the coupling between the electronic and magnetic systems. 

In interaction with the electron gas the shape of the domain wall will
change as to minimize the total energy.
To evaluate the thermodynamic potential and conductance of the electron gas
within the profile (\ref{profile}) by changing the length $\xi$,
we employ a multichannel $S$-matrix formalism\cite{s_matrix,buttiker,jellium} 
where,
\begin{equation*}
{\bf S}=
\begin{pmatrix}
S_{ll,\uparrow\uparrow}
&S_{ll,\uparrow\downarrow}
&S_{lr,\uparrow\uparrow}
&S_{lr,\uparrow\downarrow}\\
S_{ll,\downarrow\uparrow}
&S_{ll,\downarrow\downarrow}
&S_{lr,\downarrow\uparrow}
&S_{lr,\downarrow\downarrow}\\
S_{rl,\uparrow\uparrow}
&S_{rl,\uparrow\downarrow}
&S_{rr,\uparrow\uparrow}
&S_{rr,\uparrow\downarrow}\\
S_{rl,\downarrow\uparrow}
&S_{rl,\downarrow\downarrow}
&S_{rr,\downarrow\uparrow}
&S_{rr,\downarrow\downarrow},
\end{pmatrix}
\end{equation*}
\noindent

\noindent
and $l,r$ the left and right channels.

The conductance $G$ is given by,

\begin{equation}
G=\frac{e^2}{h}\int_0^{\infty} d\epsilon 
(-\frac{\partial f}{\partial \epsilon} )
tr (S^{\dagger}_{rl}S_{rl})
\label{g}
\end{equation}
\noindent
where $S_{lr},~S_{rl}$ are 2 by 2 matrices in spin space and 
the trace is over the spin indices. $f(\epsilon)$ is the Fermi function,
\begin{equation*}
f(\epsilon)=\frac{1}{1+e^{\beta(\epsilon-\mu)}}
\end{equation*} 
\noindent
with $\beta=1/k_BT$, $T$ the temperature and $\mu$ the chemical potential.
In the following we take $e^2/\hbar=1$, $\hbar^2/2m=1$ so that 
$\epsilon=q^2$, $k_B=1$ and consider the zero temperature limit, 
$T\rightarrow 0 (\beta\rightarrow \infty)$.

The electronic density of states $D(\epsilon)$ is given by,

\begin{eqnarray}
D(\epsilon)&=&\frac{1}{2\pi i} \sum_{ab} tr
(S^{\dagger}_{ab}\frac{\partial S_{ab}}{\partial \epsilon} 
-S_{ab}\frac{\partial S^{\dagger}_{ab}}{\partial \epsilon}).
\nonumber\\
&=&\sum_{a,b,\sigma,\sigma'} \frac{1}{\pi}\rho^2_{a,b,\sigma,\sigma'} 
\frac{\partial\phi_{a,b,\sigma,\sigma'}}{\partial \epsilon}
\end{eqnarray}

\noindent
with $S_{a,b,\sigma,\sigma'}=\rho_{a,b,\sigma,\sigma'} 
e^{i\phi_{a,b,\sigma,\sigma'}}$  
($a,b=l,r,~~\sigma,\sigma'=\uparrow,\downarrow$)
and the grand canonical potential by,
\begin{equation}
\Omega=-T\int_0^{\infty} d\epsilon D(\epsilon) \ln(1+e^{-\beta(\epsilon-\mu)}).
\label{energy}
\end{equation}

We construct the Bloch 
domain wall $\bf S$ matrix by decomposing the interval $L$ in slices 
of width $dx$ and by successive $x$-dependent $R_y(\theta(x))$ 
rotations which make the scattering 
diagonal in each slice. 

\begin{equation}
R_y(\theta)=
\begin{pmatrix}
+\cos(\theta/2)&+\sin(\theta/2)\\
-\sin(\theta/2)&+\cos(\theta/2)
\end{pmatrix}
\end{equation}

\noindent
For a N\'eel wall the rotation matrix is,
\begin{equation}
R_x(\theta)=
\begin{pmatrix}
+\cos(\theta/2)&+i\sin(\theta/2)\\
+i\sin(\theta/2)&+\cos(\theta/2).
\end{pmatrix}
\end{equation}

In this one dimensional model, the Bloch and N\'eel walls are related 
to each other by a rotation of the local quantization axis, 
thus giving identical thermodynamic potentials and conductances. 
Furthermore we assume $h(x)=h$ independent of position,
although it is straightforward to consider domain walls with 
position dependent coupling $h(x)$. 
We note that we verified the 
S-matrix calculation by a T-matrix method, although 
we found that the T-matrix approach is numerically unstable for large 
$L$ systems due to the appearance of exponentially large terms. 

\section{Results}
\subsection{Single wall}
\begin{figure}[ht]
\includegraphics[angle=0, width=.8\linewidth]{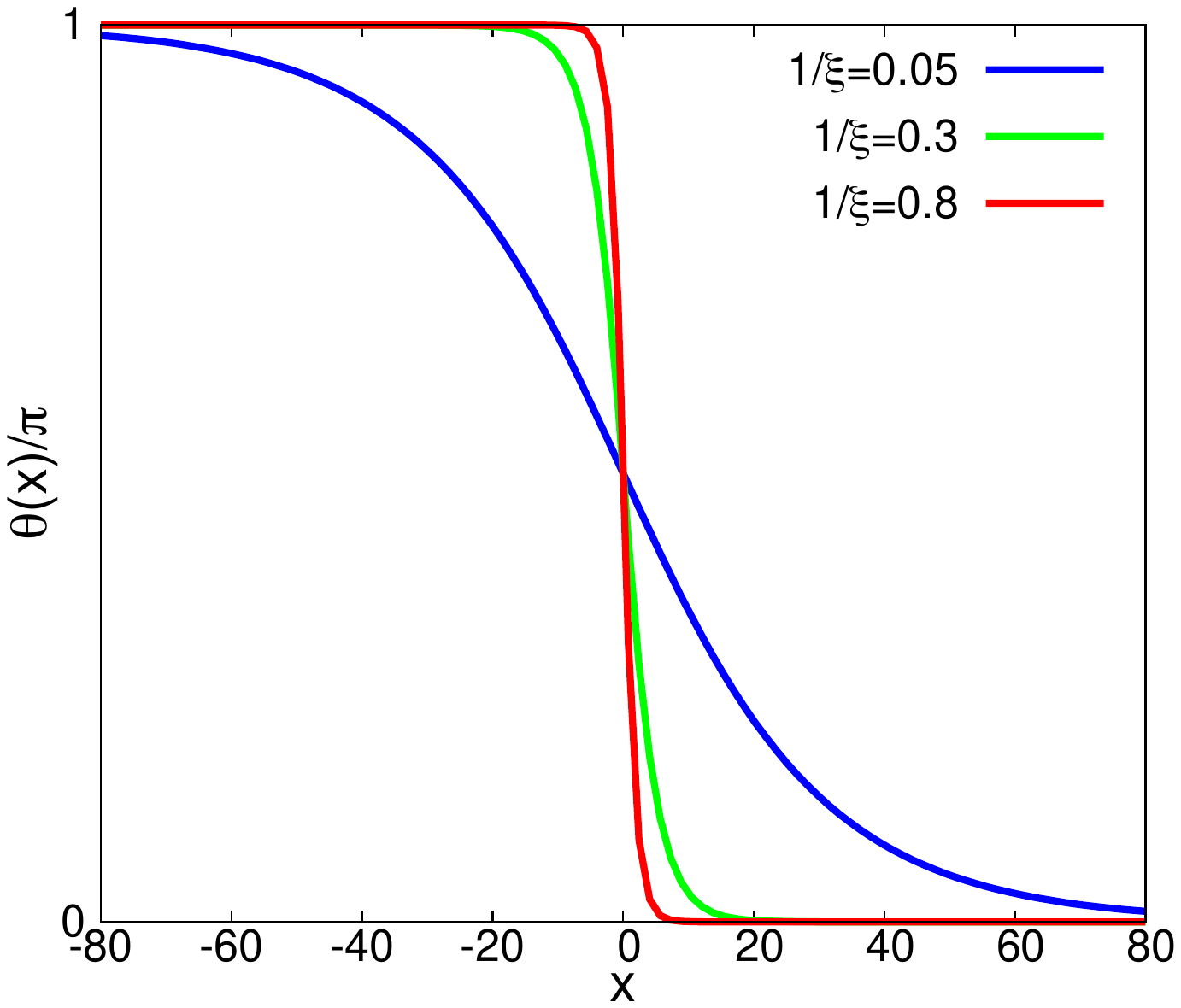}
\caption{Domain wall profile for different widths $\xi$.}
\label{wall}
\end{figure}

\begin{figure}[ht]
\includegraphics[angle=0, width=.8\linewidth]{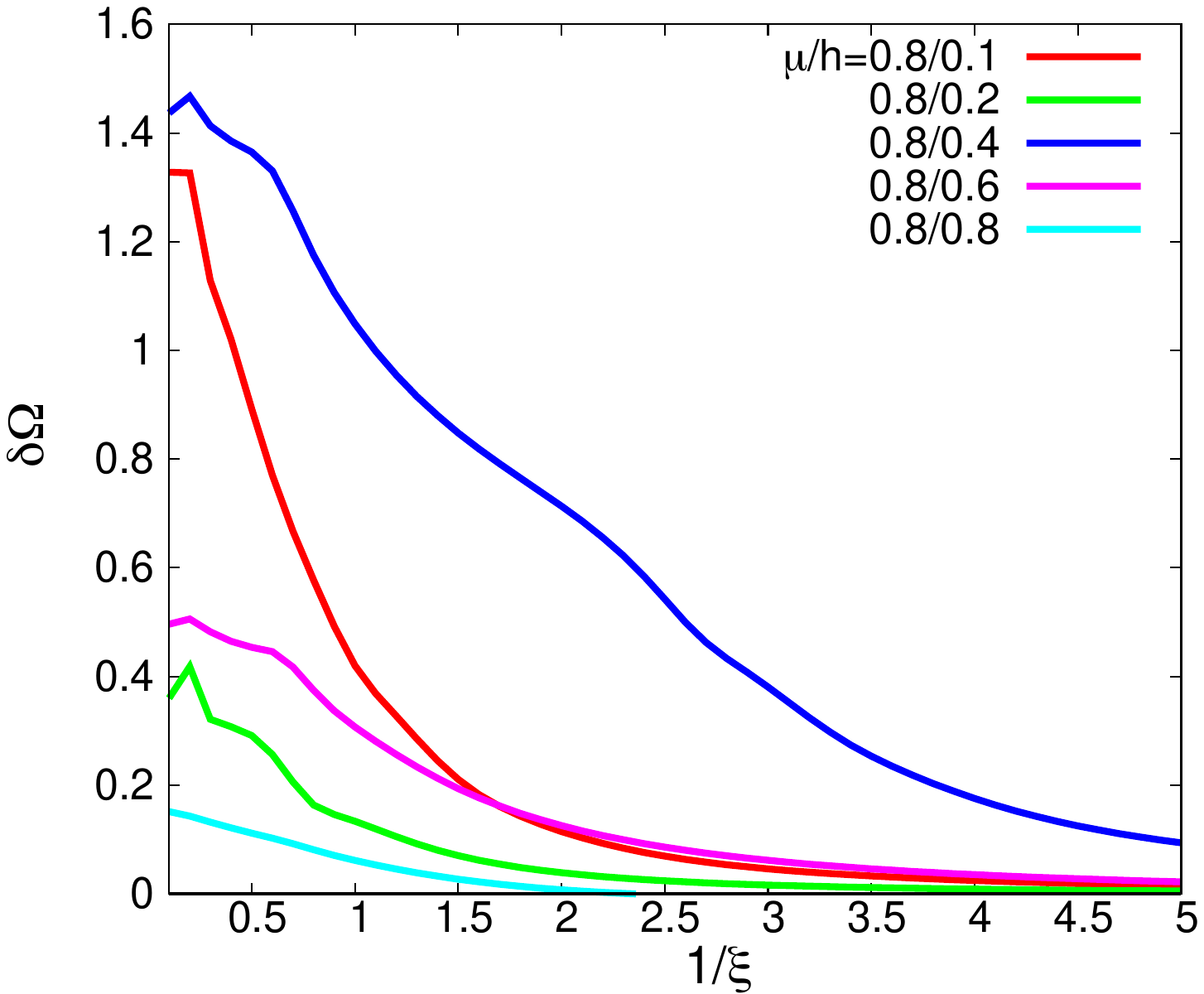}
\caption{Normalized grand canonical potential  
$\delta \Omega=\Omega - \Omega_{\xi\rightarrow 0}$ of the electronic gas 
in the presence of a magnetic domain wall for  
$\mu=0.8$ and different fields $h$.}
\label{fig2}
\end{figure}

\begin{figure}[ht]
\includegraphics[angle=0, width=.8\linewidth]{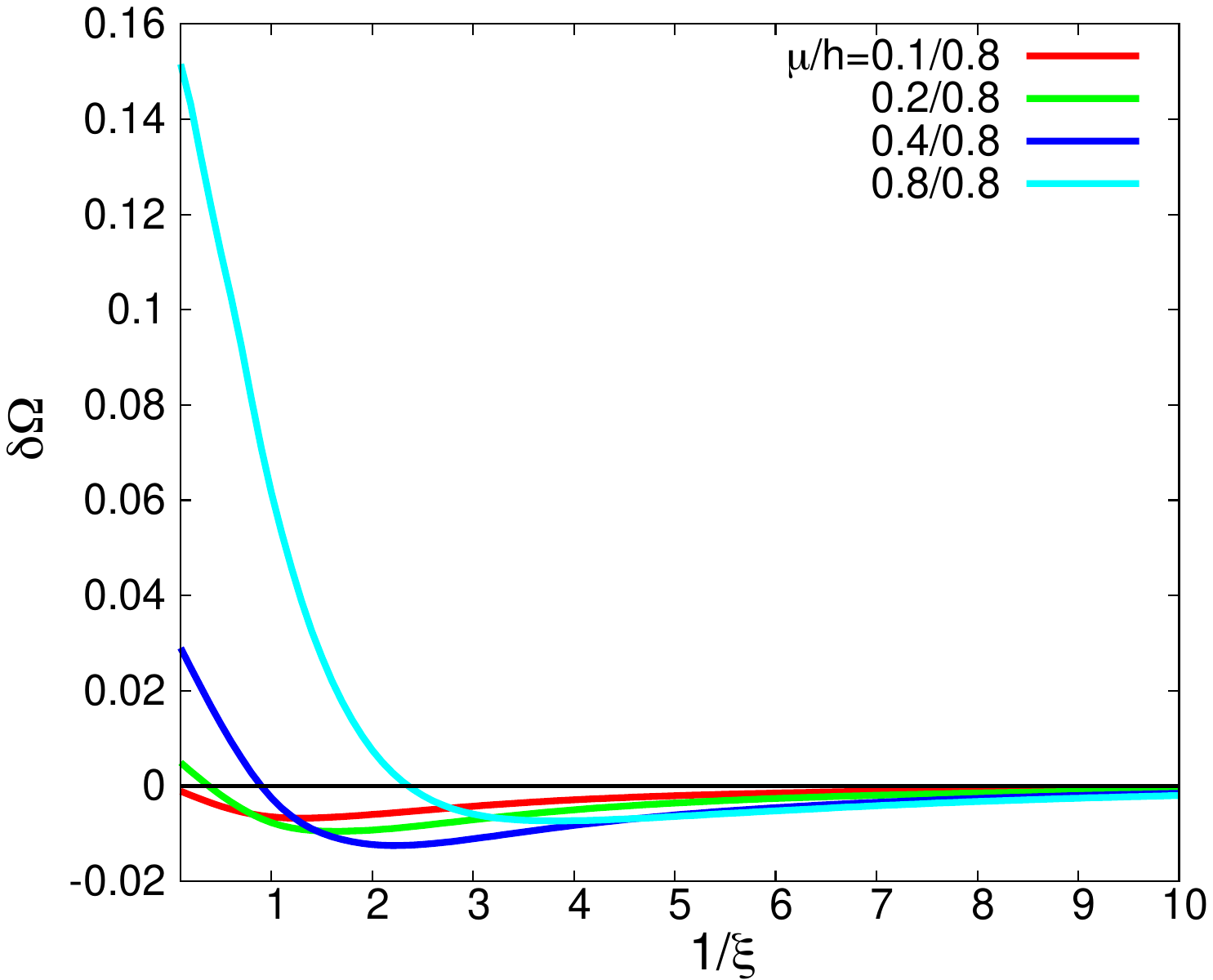}
\caption{Normalized grand canonical potential  
$\delta \Omega=\Omega - \Omega_{\xi\rightarrow 0}$ of the electronic gas 
in the presence of a magnetic domain wall for  
$h=0.8$ and different chemical potentials $\mu$.}
\label{fig3}
\end{figure}

\begin{figure}[ht]
\includegraphics[angle=0, width=.8\linewidth]{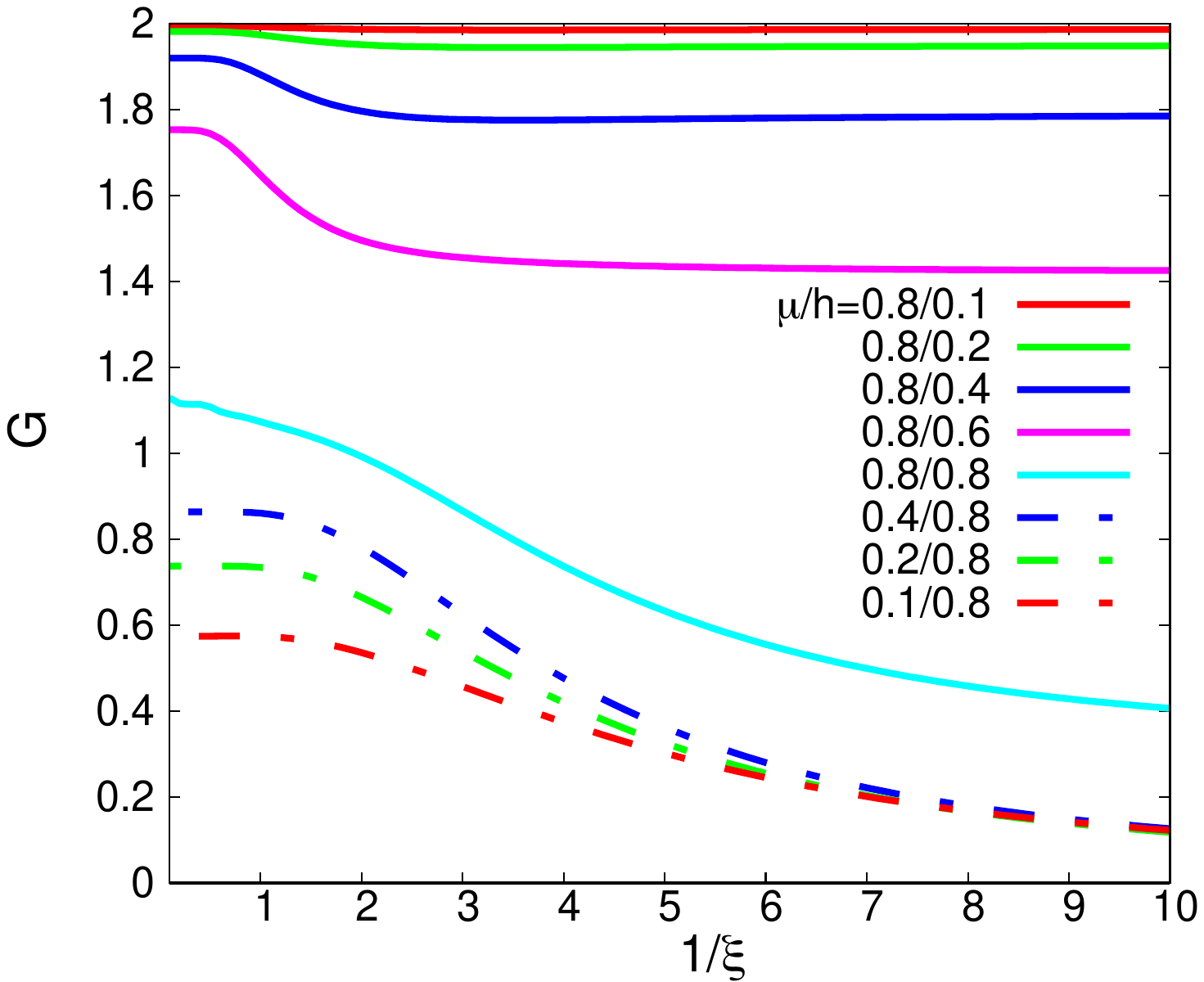}
\caption{Conductance of the electronic gas in the presence of a magnetic 
domain wall 
for different ratios of magnetic field $h$  and 
chemical potential $\mu$, $\mu/h~(\mu=0.8~ {\rm or}~ h=0.8)$.}
\label{fig4}
\end{figure}

To obtain the $S$-matrix we split the magnetic wall domain 
$L=160~(-80 < x< 80)$ in 800 slices of width dx=0.2. 
Examples of the profiles of 
domain walls we are considering are shown in Fig.\ref{wall}.
In Figs.\ref{fig2},\ref{fig3}
we show the thermodynamic potential and in Fig.\ref{fig4} the corresponding 
conductance.

We find that for $\mu > h$ the thermodynamic potential is generally minimized 
for narrow domain walls ($\xi \rightarrow 0$). We note however that there are 
exceptional $\mu/h$ values where the minimum is at a finite $\xi$, probably 
due to resonant scattering. 
In quantum mechanical coherent systems there are subtle 
dependencies of the phase shifts due to interferences, 
on the energy, size, shape and magnitude of the barriers \cite{bruno}.
In contrast, for $h > \mu $ 
the dependence of the thermodynamic potential on the width is nonmonotonic 
with minimum at $\xi$ of order one.
As expected, the conductance takes practically the non-interacting limit 
value $G=2$ for $\mu >> h$ and it is rather weakly dependent on $\xi$. 
For $\mu < h$, it is supressed to $G < 1$.
Thus we conclude that 
for $\mu > h$ the width of the  wall has a sizable effect 
on the electronic energy and a rather smooth one on the conductance.

The final width of the wall will be determined by the competition of 
elastic and electronic energy, namely the ratio of exchange $J$ to 
anisotropy interaction $D$, the chemical potential $\mu$ and magnetic coupling 
$h$. 
From the above data, we generally expect the width of 
the wall for $\mu > h$ to be drastically reduced due to the 
interaction with the electron bath.
We should note that, (i) doubling the domain wall length to $L=320$ 
gives a similar but proportional to the length $\Omega(1/\xi)$ curve 
and (ii) the search of total minimum energy could be 
extended to other magnetic wall profiles.
  
\subsection{Double domain walls}
\begin{figure}[htp]
\includegraphics[angle=0, width=1\linewidth]{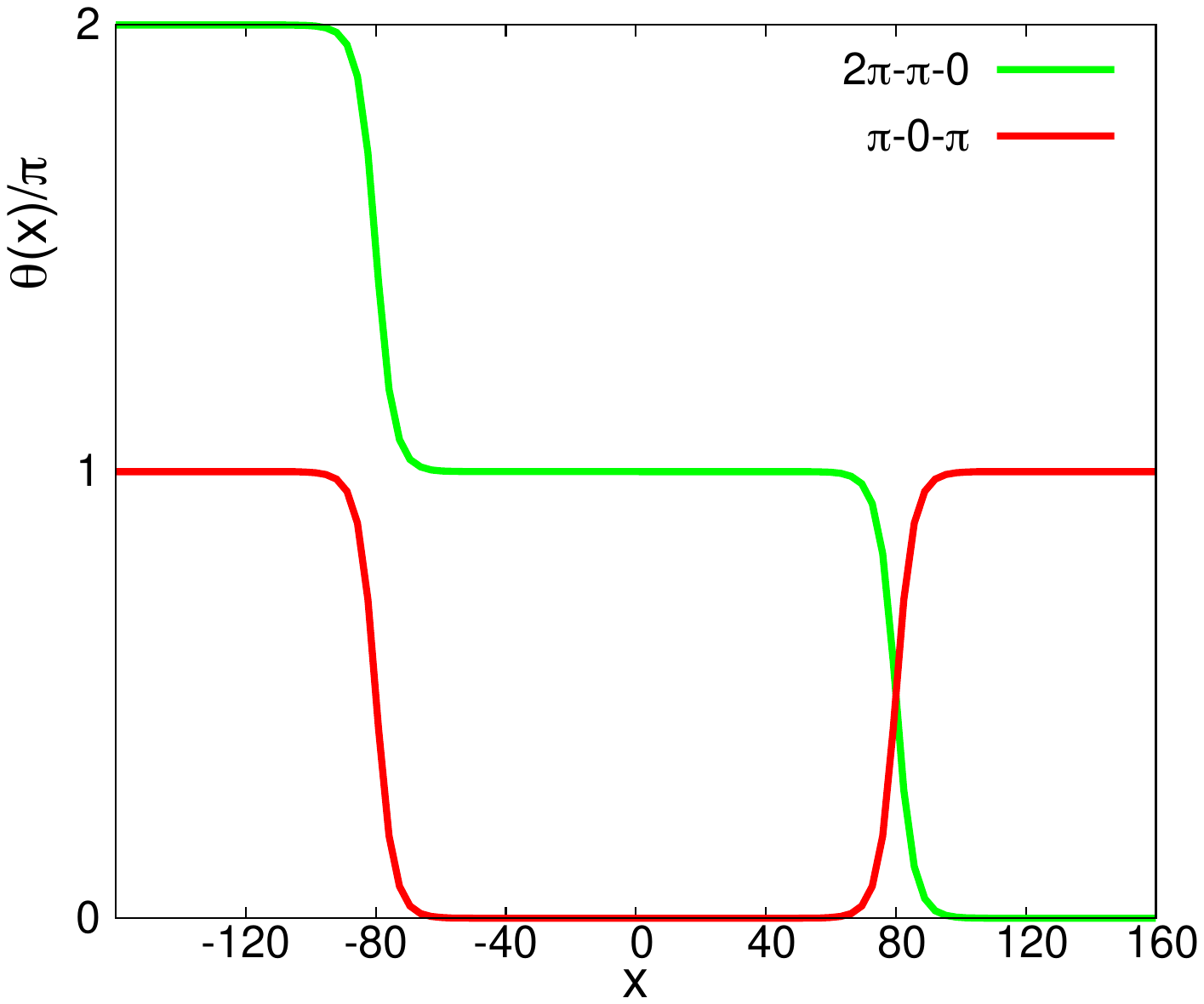}
\caption{Configurations of two adjacent domain walls with $1/\xi=0.3$.} 
\label{wall2}
\end{figure}

In this section we study the thermodynamic potential 
and conductance of two adjacent magnetic 
domain walls, in two different relative chirality 
configurations, $2\pi-\pi-0$, $\pi-0-\pi$, as shown in Fig.\ref{wall2}.
For the $2\pi-\pi-0$ domain wall,
\noindent
\begin{equation}
\theta(x)=2\tan^{-1} e^{-(x-L/4)/\xi}+2\tan^{-1}e^{-(x+L/4)/\xi},
\label{2pipi0}
\end{equation}

\noindent
and for the $\pi-0-\pi$,
\noindent
\begin{equation}
\theta(x)=2\tan^{-1} e^{+(x-L/4)/\xi}+2\tan^{-1}e^{-(x+L/4)/\xi}.
\label{pi0pi}
\end{equation}

\noindent
To obtain the $S$-matrix we split the magnetic walls region in 1600 slices 
of width dx=0.2. 

\begin{figure}[htp]
\includegraphics[angle=0, width=1\linewidth]{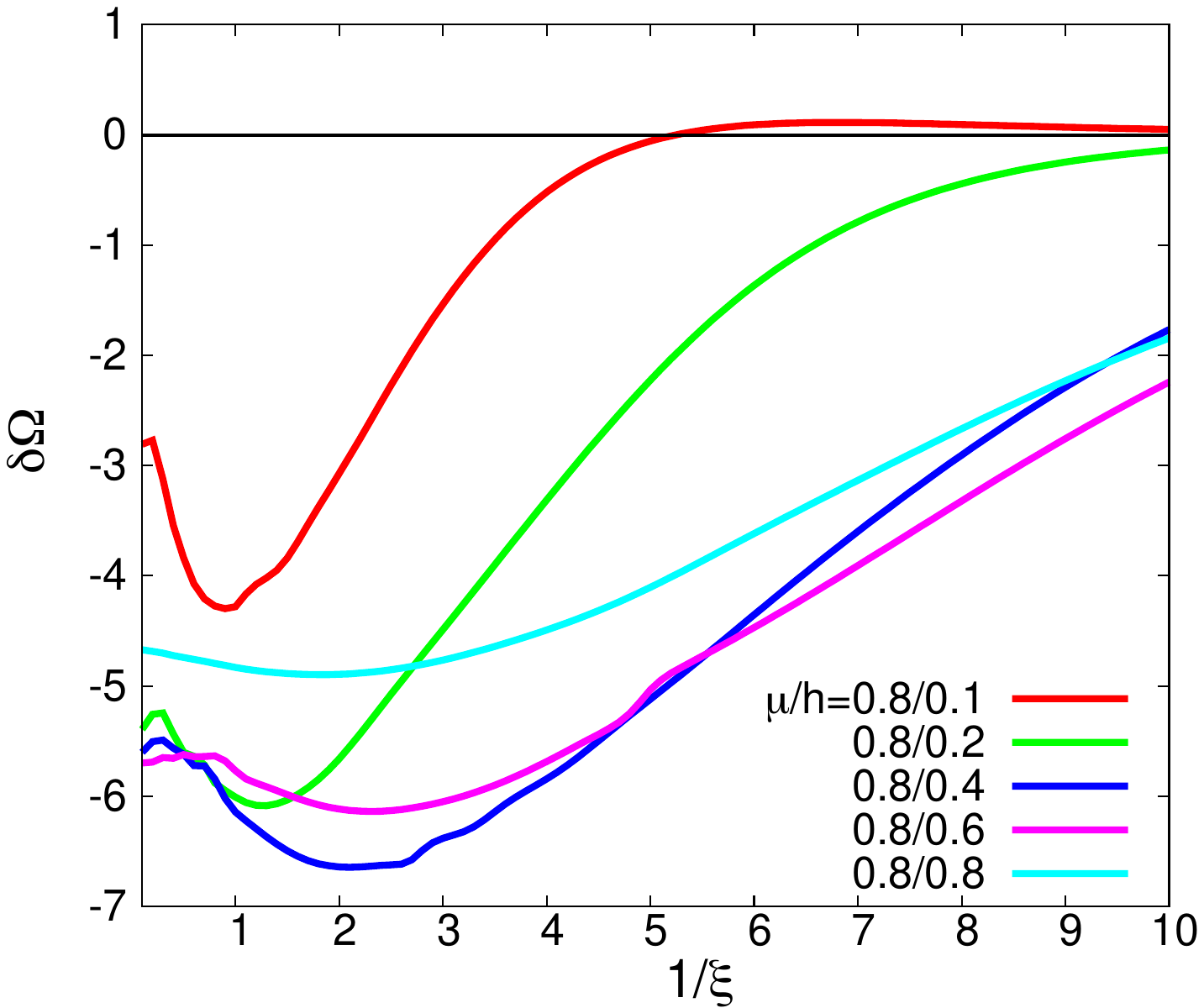}
\caption{Normalized grand canonical potential  
$\delta \Omega=\Omega - \Omega_{\xi\rightarrow 0}$ of the electronic gas 
in the presence of a double magnetic domain wall 
$2\pi-\pi-0$ or $\pi - 0 - \pi$ for  
$\mu=0.8$ and different fields $h$.}
\label{fig6}
\end{figure}

\begin{figure}[htp]
\includegraphics[angle=0, width=1\linewidth]{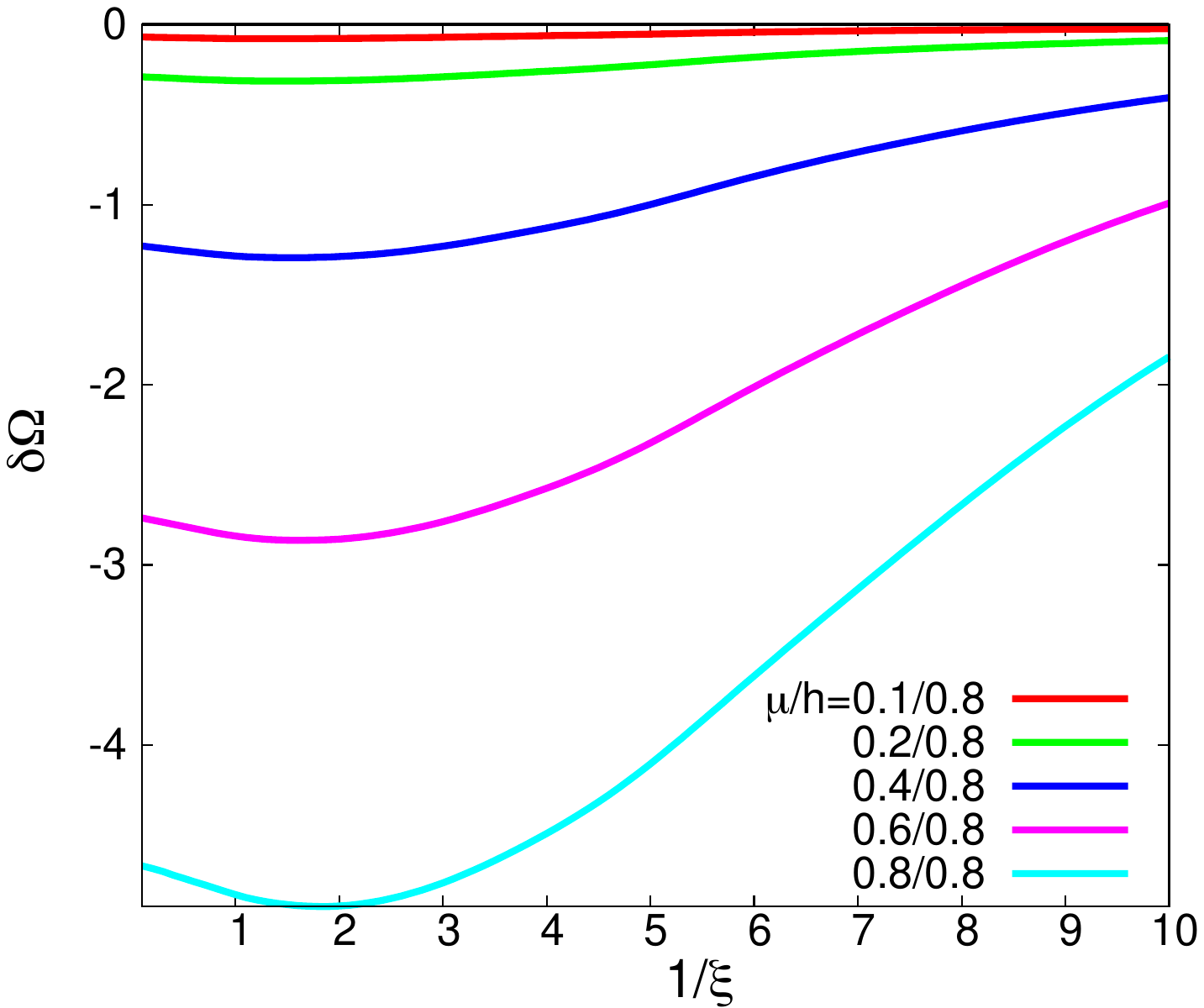}
\caption{Normalized grand canonical potential  
$\delta \Omega=\Omega - \Omega_{\xi\rightarrow 0}$ of the electronic gas 
in the presence of a double magnetic domain wall 
$2\pi-\pi-0$ or $\pi - 0 - \pi$ for  
$h=0.8$ and different chemical potentials $\mu$.}
\label{fig7}
\end{figure}

In Figs.\ref{fig6},\ref{fig7} we show the grand canonical 
potential as a function of $1/\xi$, that we find identical 
for the  $2\pi-\pi-0$  and $\pi-0-\pi$ walls. 
In general the dependence of the thermodynamic potential on the 
domain wall width is non-monotonic, with minimum of the order $\xi \sim 1$. 

\begin{figure}[htp]
\includegraphics[angle=0, width=1\linewidth]{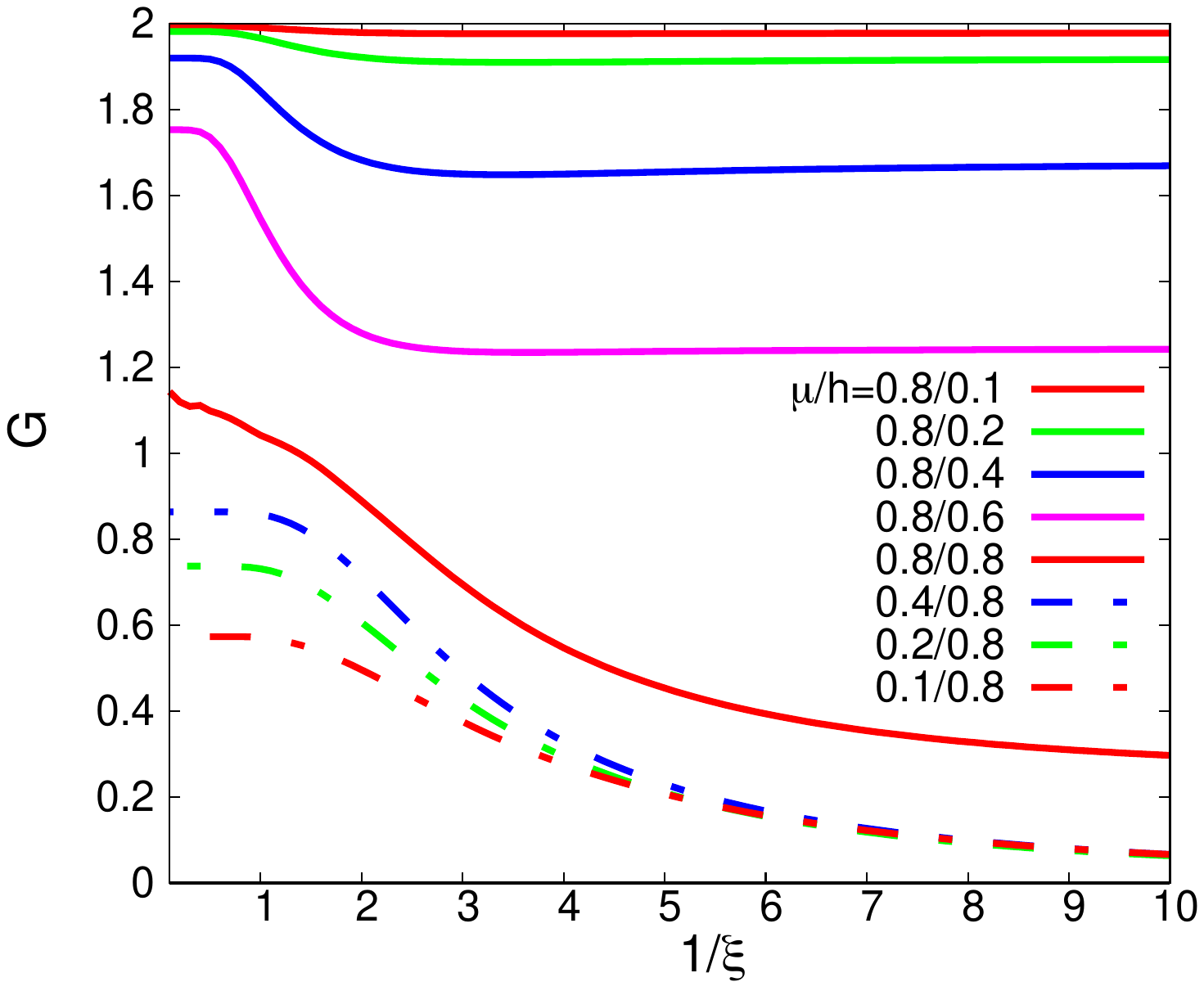}
\caption{Conductance of the electronic gas in the presence of a 
double magnetic domain wall 
for different ratios of magnetic field $h$  and 
chemical potential $\mu$, $\mu/h~ (\mu=0.8~ {\rm or}~ h=0.8)$.}
\label{fig8}
\end{figure}

Similarly, as shown in Fig.\ref{fig8}, 
the conductance is identical for the $2\pi-\pi-0$ and $\pi-0-\pi$ 
walls and qualitatively similar to the single wall conductance
Thus the thermodynamic potential and conductance of the 
double wall configurations are independent on their chirality.
Finally, at low temperatures ($T\sim 0.1$) the data for both the single as well 
as the double domain walls remain qualitatively similar to those in the 
zero temperature limit. 

\section{Conclusions}
We have studied the thermodynamics and transport of 
an open quantum mechanical coherent system using
an S-matrix formulation.
The S-matrix approach we used provides a rigorous and unified picture 
of the conductance as well as the thermodynamic potential 
of the electronic gas. 
We found that the double exchange interaction of a ferromagnetic
magnetic domain wall with an independent
electron gas causes a significant deformation of the wall
when the chemical potential is larger than the magnetic interaction coupling
(the results are consistent with perturbative ones in \cite{tatara}).
In contrast, the conductance is generally rather weakly dependent on the width 
of the domain wall. 
The study of this one dimensional prototype model 
by the S-matrix approach provides a 
generic example to the problem of deformation of magnetic textures 
due to the interaction with an electronic system. 
The quantum coherent systems studied are most relevant to mesoscopic 
physics.
Hopefully this generic effect will be studied more systematically 
in future experiments. 
This study can also be extended to other types  
and higher dimensional magnetic textures 
as for instance the multitude of skyrmion configurations. 
Last but not least, the effect of quantum fluctuations 
on the magnetic wall (
corresponding to small spin) is of course a very interesting 
bul also challenging theoretical problem.

\section{Acknowledgements}
X.Z. would like to thank C. Panagopoulos, G. Kioseoglou and 
I. Rousochatzakis 
for useful discussions on the 
state of the magnetic texture deformation problem.

\end{document}